\documentclass{article}

\usepackage{PRIMEarxiv}
\usepackage{cite}
\usepackage[utf8]{inputenc} 
\usepackage[T1]{fontenc}    
\usepackage{hyperref}       
\usepackage{url}            
\usepackage{booktabs}       
\usepackage{amsfonts}       
\usepackage{nicefrac}       
\usepackage{microtype}      
\usepackage{lipsum}
\usepackage{fancyhdr}       
\usepackage{graphicx}       
\graphicspath{{media/}}     

\title{Multi-ship cooperative air defense model based on queuing theory
\thanks{Zhongyao Ma is the first author.}
\thanks{Keyu Wu is the corresponding author.}
}

\author{
 Zhongyao Ma \\
  College of Systems Engineering\\
  National University of Defense Technology\\
  Changsha \\
  \texttt{ma\_zhongyao@nudt.edu.cn} \\
   \And
 Keyu Wu \\
  College of Systems Engineering\\
  National University of Defense Technology\\
  Changsha \\
  \texttt{keyuwu@nudt.edu.cn} \\
  \And
 Zhong Liu \\
  College of Systems Engineering\\
  National University of Defense Technology\\
  Changsha \\
  \texttt{phillipliu@263.net} \\
}

\begin{document}
\maketitle

\begin{abstract}
Multi-ship air defense is an important combat task. The study of the multi-ship air defense model is of great significance in the simulation and evaluation of the actual combat process, the demonstration of air defense tactics, and the improvement of the security of important targets. The traditional multi-ship air defense model does not consider the coordination between ships, and the model assumptions are often too simple to effectively describe the capabilities of the multi-ship cooperative air defense system in realistic combat scenarios. In response to the above problems, this paper proposes a multi-ship cooperative air defense model, which effectively integrates the attack and defense parameters of both sides such as missile launch rate, missile flight speed, missile launch direction, ship interception rate, ship interception range, and the number of ship interception fire units. Then, the cooperative interception capability among ships is modeled by the method of task assignment. Based on the queuing theory, this paper strictly deduces the penetration probability of the cooperative air defense system, and provides an analytical calculation model for the analysis and design of the cooperative air defense system. Finally, through simulation experiments in typical scenarios, this paper studies and compares the air defense capabilities of the system in two different modes with and without coordination, and verifies the superiority of the multi-ship cooperative air defense model in reducing the probability of missile penetration;  Further, the ability changes of the defense system under different parameters such as missile speed, speed, angle, ship interception rate, range, and number of fire units are studied, and the weak points of the defense formation, defense range settings, and interception settings are obtained. The analysis conclusion of this paper is instructive for the attack-defense game and system design of multi-ship cooperative air defense.
\end{abstract}


\section{Introduction}
Air defense and missile defense is an important combat task. Studying the multi-ship collaborative air defense model is useful for simulating and evaluating the actual combat process of air defense and missile defense, discovering weak links in the anti-missile system, studying air defense tactics to reduce the probability of missile penetration, enhancing the survivability of important targets, and building anti-missile system. 

There are many indicators to evaluate the defense effectiveness of an air defense model, such as the ADC evaluation model, etc., where the penetration probability represents the probability of hitting the target after the missile breaks through the defense unit and intercepts it, which can intuitively reflect the defense effectiveness of the defense unit. Therefore, this paper uses the penetration probability as the evaluation index of the quality of the model. There are a lot of queuing phenomena in life, such as passengers queuing for tickets, and the telephone line in the city is busy. Usually, queuing theory is used to study such problems. Queuing theory is widely used in all service systems, especially communication systems \cite{paper1} \cite{paper2}, traffic systems \cite{paper3}\cite{paper4}, computer storage systems \cite{paper5}, production management systems \cite{paper6}\cite{paper7}\cite{paper8} et al. In the air defense problem, queuing theory has been applied in many aspects such as force deployment optimization \cite{paper9}\cite{paper10}, unit formation \cite{paper11}\cite{paper12} and air defense effectiveness evaluation \cite{paper13}\cite{paper14}. In the context of this paper, the process of missile penetration and interception by ships is the queuing process. When the missile flow enters the interception range of the ship, it forms a queue and waits to be intercepted in turn. Therefore, the penetration probability of this model can be calculated by using queuing theory.

The purpose of this paper is to construct a set of air defense and anti-missile model with high credibility, high flexibility, wide applicability, and fitting to realistic combat scenarios, in which the synergistic cooperation among multiple interceptor ships and joint interception of incoming missiles are the basic requirements of a modern maritime air defense and anti-missile model. The U.S. Army proposed the Naval Integrated Control-Air Control System as early as 1996, with the aim of achieving a distributed, networked, multi-level maritime cooperative air defense and anti-missile \cite{paper15}. After years of development, the U.S. military is now able to achieve not only cooperative air defense among multiple ships, but also reach the intersection of sea, land, and air kill chains, which plays an important role in the construction of the kill network. In the past, Jiang Tao et al \cite{paper16} first proposed a set of air defense simulation models to evaluate the penetration probability, which can dynamically set parameters such as missile launch position and number, air defense fire unit position and number, etc., and can solve the penetration probability in a single air defense area as well as evaluate the penetration probability under the joint air defense of multiple air defense areas, establishing a prototype of the synergistic air defense model, but it has several shortcomings: (1) The cooperative air defense strategy and the derivation process of the overall penetration probability of the system in the cooperative air defense scenario are not given. (2) The adjustable parameters of the model are relatively rough, lacking real parameters in the real world such as missile flight speed, launch rate, defense ship interception rate, and interception range. (3) There is a lack of comparative experiments between the cooperative model and the non-cooperative model to prove the superiority of the model cooperative air defense. (4) It is relatively simple to discuss the practical significance of the model through the experimental results.

In response to the above problems, this paper proposes a multi-ship cooperative air defense model. This paper is structured as follows: firstly, the multi-ship cooperative air defense model is modeled in five aspects: incoming missiles, interceptor ships, queuing rules, collaborative model and task assignment rules. Then each interceptor ship is considered as an M/M/m queuing model, and the queuing theory is used to derive an expression for the penetration probability for the model as a whole. Finally, the comparative experiments of the cooperative model and the non-cooperative model are carried out, and the superiority of cooperative air defense is verified by comparing the penetration probability of the two under the same conditions. At the same time, the guiding significance of the model results for practical application is obtained by analyzing the variation curve of model capability under different parameters.

\section{Problem Modeling}
The problem studied in this paper is to improve the traditional model without considering the cooperative interception between ships, and propose a multi-ship cooperative air defense model, which aims to enhance the air defense capability of the model and reduce the penetration probability of missiles. The penetration probability is defined as the probability that the incoming missile will not be intercepted by the ship and hit the target during the flight to the target after entering the interception range of the ship. Let the flight time of the missile from entering the interception range to striking to defending the target is $W_t$, the waiting time of the missile from entering the interception range to being intercepted by the ship is $W_T$. According to the definition of penetration probability, the calculation of penetration probability P is shown in the following Eq. (1).
\begin{equation}
P=P[W_T>W_t]
\end{equation}
Each ship in this model can be regarded as a queuing system, in which the ship is the service desk, the incoming missile is the customer, the process of the missile entering the interception range and being intercepted is the service process, and the time required for the ship to intercept the missile is the service time. Therefore, this paper deduces the penetration probability of the whole model based on queuing theory.

Let our defense target position loc0 be the origin (0,0).The adversary uses missiles to conduct raids on our important defense targets from different directions. The strike task constitutes a queue, and the missiles in different directions constitute multiple queues. A strike mission means that the missiles are continuously launched to strike the target at certain time intervals. Around the defense target, a number of defense ships are arranged. When the incoming missiles fly into the interception range of the ships, the ships will intercept them in sequence according to the queuing rules. Each intercepting ship forms a coordinated defense system, and assigns missile interception tasks to different ships to maximize interception efficiency. Therefore, the model is explained from the following five aspects: incoming missiles, defensive ships, queuing rules, coordination models and task allocation rules.

\subsection{Incoming Missile}
Assuming that the same type of missiles attack the defense target from $N_T$ different directions, the missile launch position is $loc_i$. $loc_i$ and the defense target position (0,0) determines the launch direction. The missile launch method is that it is launched continuously at a certain time interval from different positions according to the strike mission. Assuming that the successive arrival time $t_i$ of the missile reaching the interception range of the ship obeys the negative exponential distribution with the parameter $\lambda_i$, the expression is shown in the following Eq. (2). The arrival process of the missile is smooth and has no aftereffect. $\lambda_i$ is the missile launch rate, indicating the speed of the missile launch, and the flight speed of the missile is $v_i$, where i=1,2,…,$N_T$.
\begin{equation}
p(t_i)=\lambda_i e^{-\lambda_i t_i}
\end{equation}

\subsection{Defensive Ship}
Assume that $N_C$ defense ships are deployed around the defense target according to a certain formation, and the ship position is $loc_j$. Assuming that the ship's interception range is a circle with radius $R_j$, different types of ships take different interception times to intercept incoming missiles. Assuming that the time interval $t_j$ required by the ship from successfully intercepting a missile to continuing to intercept the next missile obeys the negative exponential distribution with the parameter $\mu_j$, the expression is shown in the following Eq. (3). Assuming that the time intervals of ship interception are independent of each other, $\mu_j$ is the interception rate of the ship, indicating the speed at which the ship intercepts the missile. Different ships are equipped with different numbers of firepower units $m_j$, which represent the number of missiles that a ship can intercept at the same time. 
\begin{equation}
p(t_j)=\lambda_j e^{-\lambda_j t_j}
\end{equation}

\subsection{Queuing Rules}
The flow of incoming missiles from all directions reaches the ship's interception range one by one, and missiles from different directions form a queue according to the strike mission, and the queue is infinitely long. If the ship has no missiles being intercepted, the newly-flying missiles will be intercepted immediately, otherwise the missiles will join the queue and continue to fly towards the target until the previous missile interception is completed. Ships intercept incoming missiles on a first-come, first-served (FCFS) basis, and will only intercept missiles that come into interception range.

\subsection{Collaborative Model}
As the missile strikes a defended target, it passes through the interception range of multiple ships. In the traditional non-coordinated model, different ships intercept incoming missiles from a specific direction according to certain principles, which is a "one-to-one" mode of interception. In the multi-ship collaboration model proposed in this paper, the ship can intercept missiles in any direction passing through its interception range, which is a "many-to-many" mode of interception. The difference is shown in Figure 1. The left picture in the figure is a non-coordinated model, which means that the missile in direction i is completely intercepted by the corresponding ship i, and $\lambda_i$ represents the launch rate of the missile. The figure on the right shows a collaborative model, which means that the missiles in the i-direction are jointly intercepted by different ships j, j=1,2,…,$N_C$, and $\lambda_{ij}$ represents the launch rate of the missile strike task in the i-direction assigned to the interception by the ship j. Capital letters $C_n$ etc. indicate fire units on different ships.
\begin{figure}
 \centering
 \includegraphics[height=7.5cm]{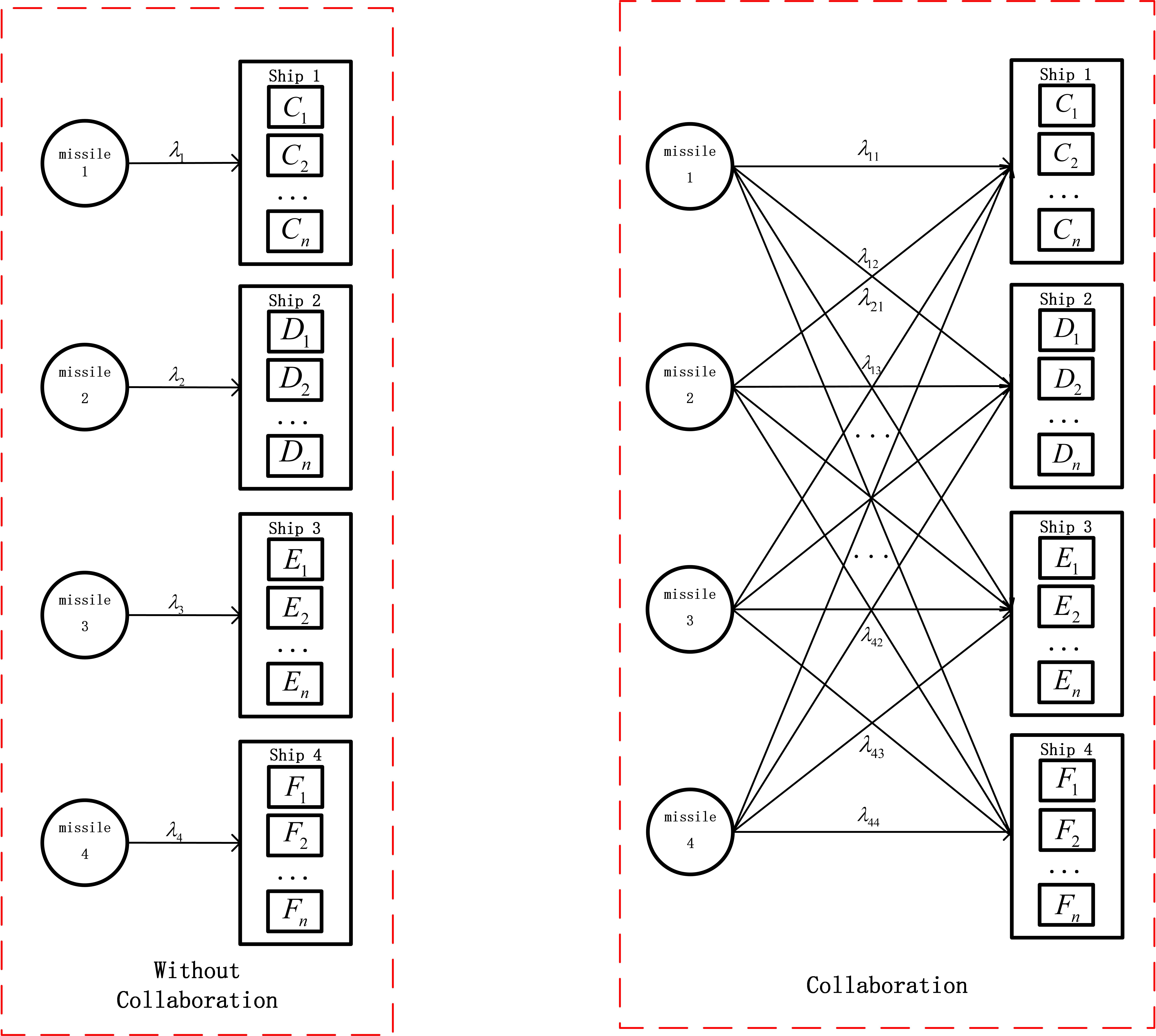}
 \caption{Comparison of models with/without Collaboration.}
 \label{fig:time}
\end{figure}

\subsection{Task Assignment Rules}
The cooperative air defense among ships is realized by the assignment of intercepting tasks to incoming missiles. In the past, only a specific ship was responsible for the strike task of missiles in a certain direction. Now the strike task is assigned to multiple ships to intercept at the same time, thereby maximizing the interception efficiency of the model. The rules for the assignment of interception tasks are set such that the greater the number of missiles traversing the interception range of a ship, the heavier the interception task assigned to the ship. Since the missiles are continuously launched according to the negative exponential distribution time interval $t_i$, the number of missiles passing through the interception range of the ship is defined by the length of the intersection of the missile trajectory and the interception range. The longer the intersection is, the more missiles enter the interception range. The intersection length of the missile trajectory in the i-direction and the interception range of the ship j is recorded as $line_{ij}$. Therefore, the firing rate $\lambda_{ij}$ after the missile attack task in the i direction is assigned to the interception of the ship j is shown in the following Eq. (4), where i=1,2,…,$N_T$,    j=1,2,…,$N_C$. 
\begin{equation}
\lambda_{ij}=\lambda_i * {\frac{line_{ij}}{\sum_j line_{ij}}}
\end{equation}
The schematic diagram of the multi-ship cooperative air defense model is shown in Figure 2. Set the number of missile attack directions $N_T$=3 and the number of ships $N_C$=3

\begin{figure}
 \centering
 \includegraphics[width=10cm]{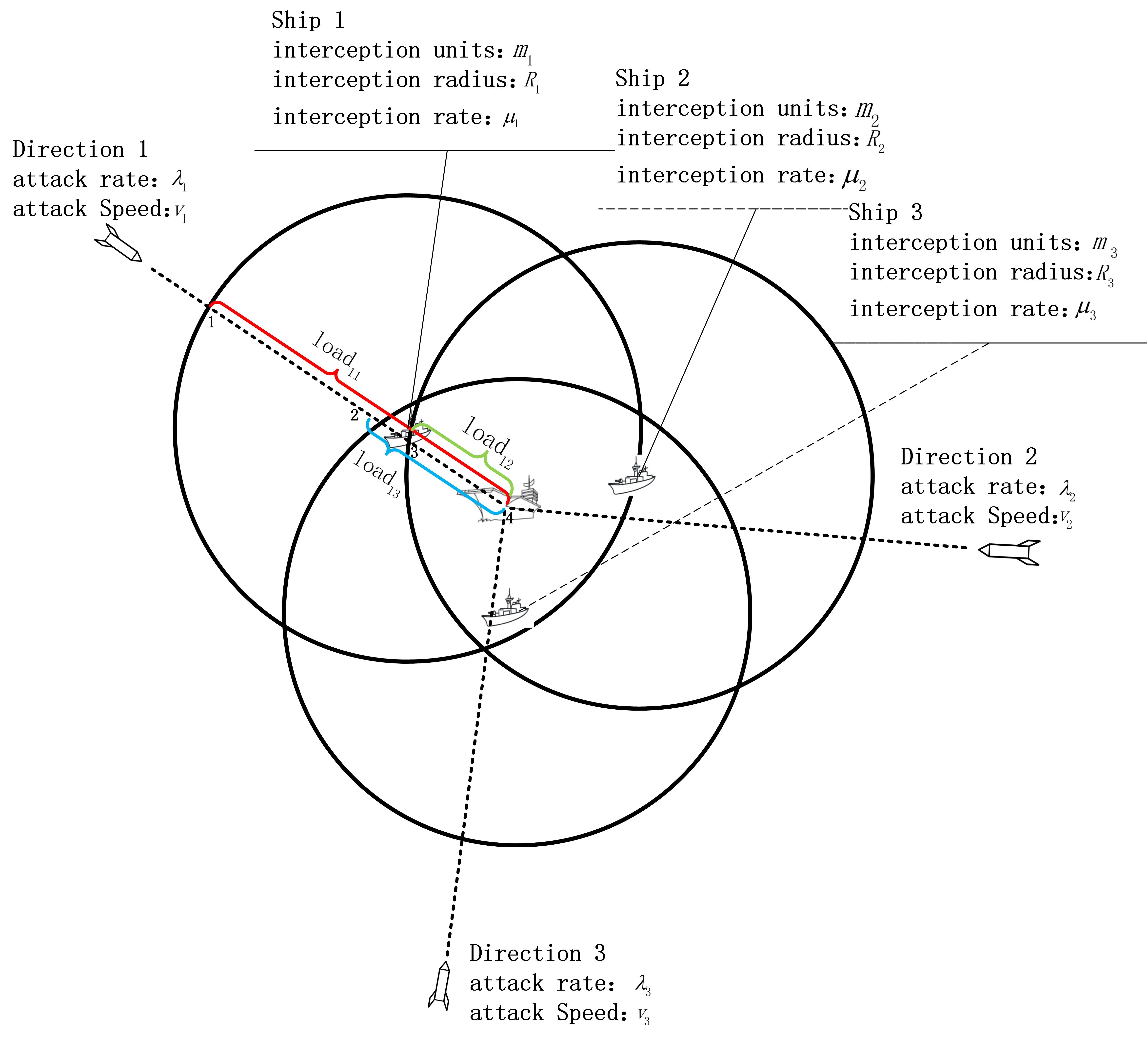}
 \caption{Multi-ship cooperative air defense model.}
 \label{fig:time}
\end{figure}

\section{Derivation of Penetration Probability}
According to the assumption in the problem modeling, the successive arrival time $t_i$ of the missile and the interception time $t_j$ of the ship obey the negative exponential distribution, so the ship can be regarded as the M/M/m queuing model, as shown in Figure 3.
\begin{figure}
 \centering
 \includegraphics[width=10cm]{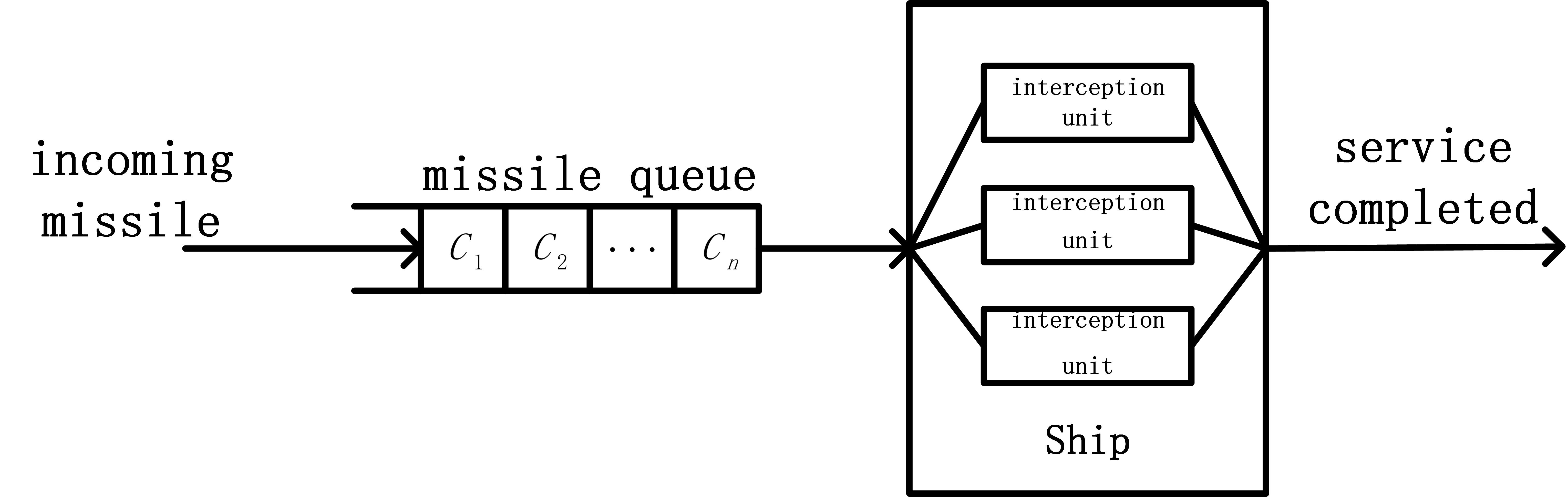}
 \caption{M/M/m queuing model of a ship.}
 \label{fig:time}
\end{figure}

\subsection{Derivation of Penetration Probability of Single Ship}
Assuming that there is only one firepower unit on ship j, the ship is in the M/M/1 queuing model. Assuming that a missile coming from a certain direction i with a launch rate $\lambda_i$ enters the interception range of the ship and finds that there are already n missiles in it, one missile is being intercepted, and the other n-1 missiles are in the queue, then the missile is waiting to be intercepted. The interception time $W_T$ is the Eq. (5). 

\begin{equation}
W_T=R_1+S_2+...+S_n
\end{equation}

Among them, $R_1$ is the remaining service time of the missile being intercepted, and $S_2$…$S_n$ are the interception time of other missiles in the queue. According to the assumption in the problem modeling, the n random variables are independent and identically distributed, and obey the negative exponential distribution with the parameter $\mu_j$, where $\mu_j$ is the interception rate of the ship. At this time, the conditional distribution of $W_T$ can be given by the n-order Erlang distribution , which is Eq. (6). 

\begin{equation}
F_{W_T}(x|n)=1-e^{-\mu_j x} \sum_{i=0}^{n-1}{\frac{(\mu_j x)^i}{i!}},x \ge 0
\end{equation}

Therefore, the probability density of $F_{W_T} (x|n)$ is Eq. (7). 

\begin{equation}
f_{W_T}(x|n)={\frac{(\mu_j x)^{n-1}}{(n-1)!}} \mu_j e^{-\mu_j x},x \ge 0
\end{equation}

Let $a_n$ be the probability distribution of the number of missiles already in the system observed by a newly arrived missile, and $\pi_n$ be the probability distribution of the number of missiles in the system when the time is long enough to reach a stable state, which is the birth and death process, and the expression is Eq. (8), Eq. (9). 

\begin{equation}
\pi_n=(1-\rho_{ij}) \rho_{ij}^n , n=0,1,2,... ; \rho_{ij}<1
\end{equation}

\begin{equation}
\rho_{ij}={\frac{\lambda_i}{\mu_j}}
\end{equation}

Among them, $\rho_{ij}$ is the ship utilization rate, which reflects the busyness of ship j intercepting missiles in direction i. By the PASTA  property, $a_n$ is the same as $\pi_n$. Therefore, the distribution of the missile waiting time $W_T$ to be intercepted is Eq. (10), Eq. (11).

\begin{equation}
F_{W_T}(x)=\sum_{n=0}^{\infty} a_n F_{W_T}(x|n)=\sum_{n=0}^{\infty} \pi_n F_{W_T}(x|n)
\end{equation}

\begin{equation}
F_{W_T}(x)=(1-\rho_{ij})F_{W_T}(x|0)+(1-\rho_{ij})\sum_{n=1}^{\infty} \rho_{ij}^n [1-e^{-\mu_j x} \sum_{i=0}^{n-1} {\frac{(\mu_j x)^i}{i!}}]
\end{equation}

Simplify the above Eq. to get Eq. (12). 

\begin{equation}
F_{W_T}(x)=P(W_T \le x)=1-\rho_{ij} e^{-\mu_j (1-\rho_{ij}) x},x \ge 0
\end{equation}

In this model, ship j actually has $m_j$ firepower units, and the ship is the M/M/$m_j$ queuing model at this time. Denote a=$\lambda_i$/$\mu_j$, and the expression of ship utilization ratio $\rho_{ij}$ is Eq. (13). From the ErlangB formula and ErlangC formula, they are Eq. (14) and Eq. (15) respectively. By generalizing Eq. (12), the distribution function of waiting time $W_T$ under the M/M/$m_j$ queuing model can be obtained as Eq. (16).

\begin{equation}
\rho_{ij}={\frac{a}{m_j}}={\frac{\lambda_i}{m_j \mu_j}}
\end{equation}

\begin{equation}
B(m_j,a_j)={\frac{{\frac{a^{m_j}}{m_j !}}}{\sum_{i=0}^{m_j} {\frac{a^i}{i!}}}}
\end{equation}

\begin{equation}
C(m_j,a)={\frac{m_j B(m_j,a)}{m_j - a(1-B(m_j,a))}}
\end{equation}

\begin{equation}
F_{W_T}(x)=1-C(m_j,a)e^{-m_j \mu_j (1-\rho_{ij}x)}=P(W_T \le x)
\end{equation}

where x is a random variable of the missile waiting time to be intercepted. Therefore, the probability that the missile waiting time exceeds x is Eq. (17).

\begin{equation}
F_{W_T}(x)=C(m_j,a)e^{-m_j \mu_j (1-\rho_{ij}x)}=P(W_T > x)
\end{equation}

According to the definition formula (1) of the penetration probability, the probability $P_{ij}$ that the ship j is penetrated by the missile in the direction i is Eq. (18) and Eq. (19). where $line_{ij}$ represents the intersection length of the missile trajectory in the i direction and the interception range of ship j, which can be calculated from the missile trajectory and the interception radius $R_j$, and $v_i$ is the flight speed of the missile.

\begin{equation}
P_{ij}=F_{W_T}^c (W_{t_{ij}})
\end{equation}

\begin{equation}
W_{t_{ij}}={\frac{line_{ij}}{v_i}}
\end{equation}

\subsection{Derivation of the Overall Penetration Probability of the Model}

For each missile's incoming direction i, first calculate the $line_{ij}$ of the connection between the missile position $loc_i$ and the defense target position $loc_0$ and the interception range of each interceptor ship's position $loc_j$, and judge the ships involved in intercepting the incoming missile in direction i according to the intersection point. Then calculate the launch rate $\lambda_{ij}$ of the missile in direction i after task assignment on different ships j, and the expression is the same as formula (4). Finally, calculate the flight time $W_{t_{ij}}$ required to hit the defense target when the missile in the i-th direction passes through the interception range of the j-th ship, and the expression is the same as Eq. (19). 

For each ship j, substitute the launch rate $\lambda_{ij}$, the ship interception rate $\mu_j$, the number of fire units $m_j$, and the flight time $W_{t_{ij}}$ into the Eq. (18) after the assignment of the incoming missile tasks in each direction, and then the missile penetration ship in the i direction can be obtained. The probability of j is Eq. (20). 

\begin{equation}
P_{ij}=F_{W_T}^c (W_{t_{ij}})=C(m_j,a_{ij})e^{-m_j \mu_j (1-\rho_{ij}) W_{t_{ij}}}
\end{equation}

Therefore, the probability of ship j being penetrated is the sum of the penetration probabilities of missiles in all directions, which is Eq. (21).

\begin{equation}
P_j=\sum_i P_{ij}
\end{equation}

For the model as a whole, as long as one ship is penetrated, the defense target is hit, that is, the model as a whole is penetrated. Therefore, the expression of the penetration probability P of the whole model is Eq. (22). 

\begin{equation}
P = \max_j P_j
\end{equation}

\section{Experiment Analysis}
This part solves two problems through simulation experiments: First, under the same conditions, compare the penetration probability of the air defense model in the two modes with and without coordination, and verify the superiority of the multi-ship cooperative air defense model; The second is to explore the changes in the defense capabilities of the air defense model under different parameters of missile speed, speed, angle, ship interception rate, range, and number of fire units, and draw conclusions about the weak points of defense formation, defense range settings, interception settings, etc.

\subsection{Model Superiority Verification}

There are 4 incoming missiles $T_i$ in different directions and 4 defense ships $C_i$, i=1,2,3,4. The defense target location is loc0(0,0). Other parameters of the model are fixed, and experiments 1 and 2 are used to explore the comparison of penetration probability in the two modes with and without coordination when only the missile launch rate or missile flight speed is changed. The experimental parameters are shown in Table 1 and Table 2, and the schematic diagram is shown in Figure 4. 

\begin{table}
 \caption{Experimental parameters}
  \centering
  \begin{tabular}{cccc}
    \toprule
    Experiment number & missile launch rate (1/s) & missile flight speed (m/s) & Other parameters \\
    \midrule
         1            &   1/(8.00-10.00)          &         3*340              &      fixed      \\
         2            &      1/9.00               &    (1.00-10.00)*340        &      fixed       \\
    \bottomrule
  \end{tabular}
\end{table}

\begin{table}
 \caption{Fixed other parameter settings}
  \centering
  \begin{tabular}{cc|ccccc}
    \toprule
    Missile number & Missile position  & Ship number & Ship location  & Intercept range & interception rate & Fire units number\\
    \midrule
         $T_1$ & (-800,0) & $C_1$  & (-200,0) & 400 & 1/60 & 7    \\
         $T_2$ & (0,800) & $C_2$  & (0,200) & 400 & 1/60 & 7    \\
         $T_3$ & (0,-800) & $C_3$  & (0,-200) & 400 & 1/60 & 7    \\
         $T_4$ & (800,0) & $C_4$  & (200,0) & 400 & 1/60 & 7    \\
    \bottomrule
  \end{tabular}
\end{table}

\begin{figure}
 \centering
 \includegraphics[height=7.5cm]{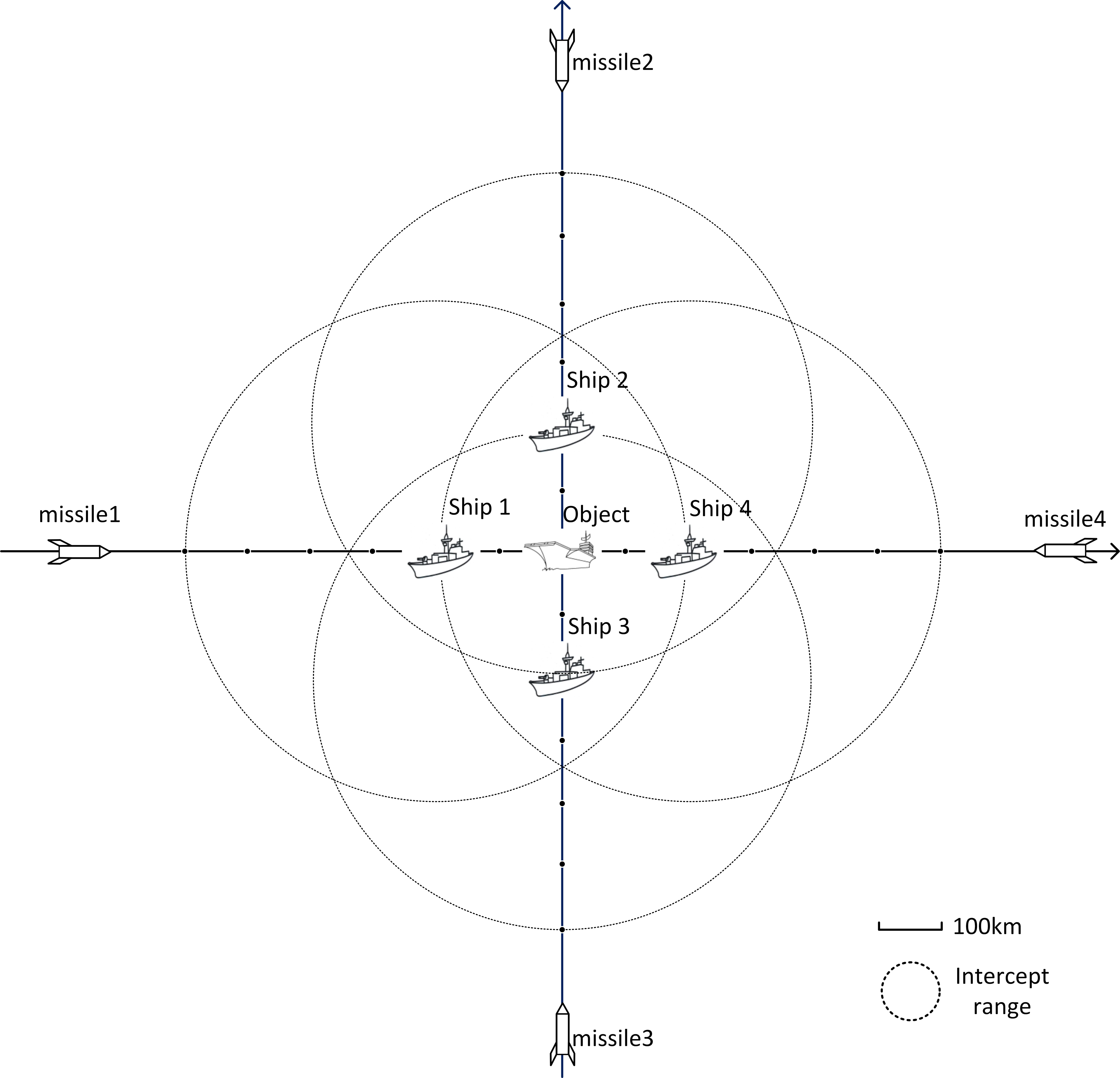}
 \caption{Schematic diagram of the experiment.}
 \label{fig:time}
\end{figure}

When the parameters of missile launch rate or missile flight speed change, the comparison of penetration probability with and without the cooperative model is shown in Fig. 5(a) and Fig. 5(b), respectively.

\begin{figure}
 \centering
 \includegraphics[width=17cm]{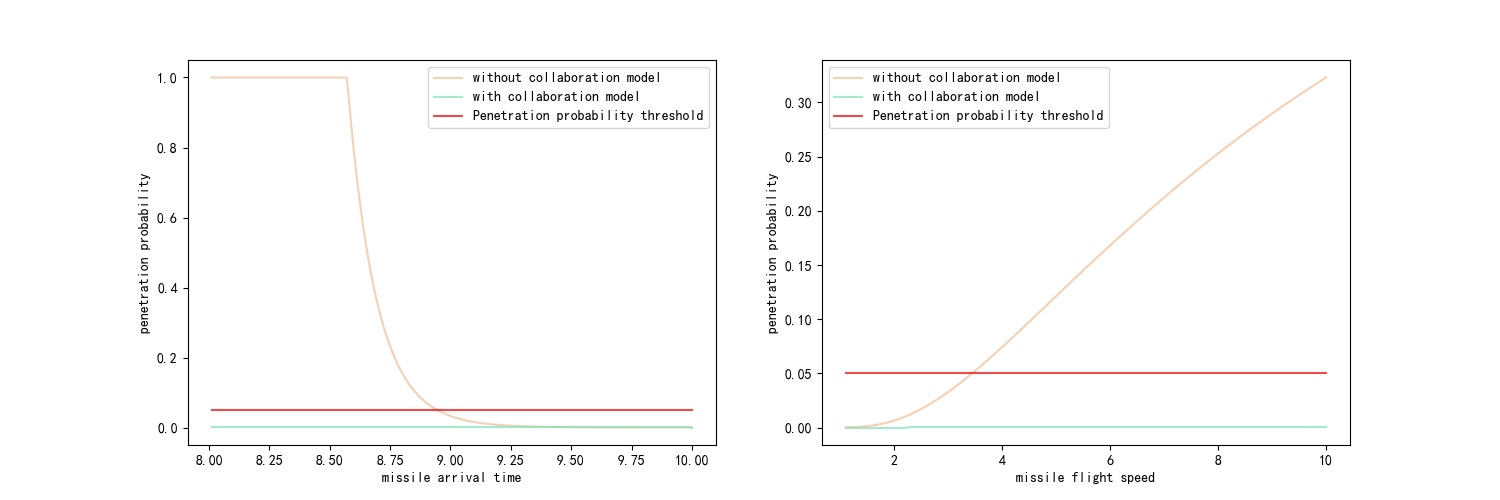}
 \caption{Comparison of penetration probability with/without collaboration model.}
 \label{fig:time}
\end{figure}

The parameters of the model without coordination will remain unchanged. In the parameters of the model with coordination, the number of firepower units will be reduced from 7 to 4, and the interception rate of ships will be reduced from 1/60 to 1/88. At this time, the comparison of penetration probability with/without collaboration model is shown in Fig. 6.

\begin{figure}
 \centering
 \includegraphics[width=17cm]{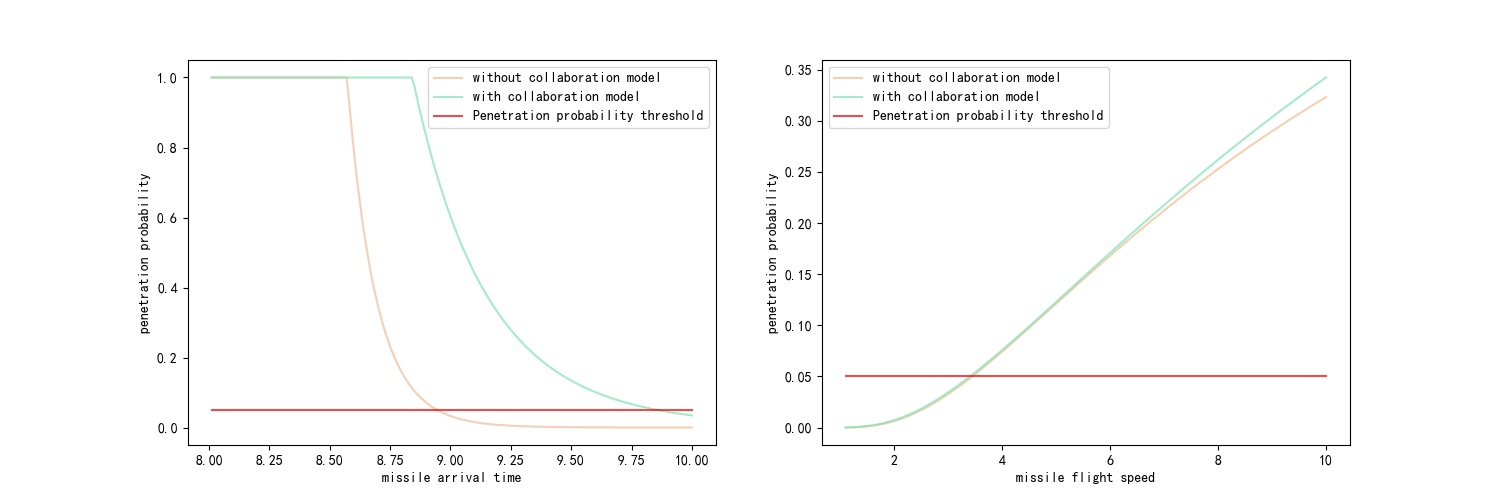}
 \caption{Comparison of penetration probability when ship parameters with synergistic model are weakened.}
 \label{fig:time}
\end{figure}

In a statistical sense, an event with a probability of less than 5 \% is generally defined as a rare event. From the perspective of collaborative air defense design, it is always hoped that the penetration of the system is a rare event, so 5 \% is uniformly set as the penetration threshold in the experiment, which can be adjusted according to needs in practical applications.

In Fig. 6(a) and Fig. 7(a), the horizontal axis is the parameter of the missile arrival time interval, which is the reciprocal of the missile launch rate. The missile arrival interval time obeys the negative exponential distribution under this parameter. The results in Fig. 6 show that, with other parameters being the same, as the missile launch rate or flight speed changes, the penetration probability of the model without coordination is significantly smaller than that of the model with coordination. The results in Fig. 7 show that under the similar defense performance of the model, the cooperative model has lower parameter requirements than the non-cooperative model, and only needs fewer firepower units and lower ship interception rate to achieve the non-cooperative model. Defensive effectiveness. The above results prove that the multi-ship air defense model with collaboration is superior to the air defense model without collaboration.

\subsection{Exploration of Model Defense Capability}

This part of the experiment explores the influencing factors of the penetration probability from the defensive end and the offensive end, and explores the guiding significance of the model in practical application by drawing the relationship between different model parameters and the penetration probability.

\subsubsection{Relationship between ship parameters and penetration probability}

This part explores the relationship between the ship parameters on the defense side and the penetration probability through the control variable method. Fix other parameters, change only a certain ship parameter, and draw the curve of the parameter changing with the penetration probability. The experimental results are shown in Fig. 7. Among them, Fig. 7(a), Fig. 7(b), Fig. 7(c) are the curves of the penetration probability with the ship's interception rate, interception range, and the number of intercepting units, respectively; Fig. 7(d), Fig. 7(e), Fig. 7(f) is the curve cluster corresponding to the above curve when the missile launch rate parameter is changed.

\begin{figure}
 \centering
 \includegraphics[width=17cm]{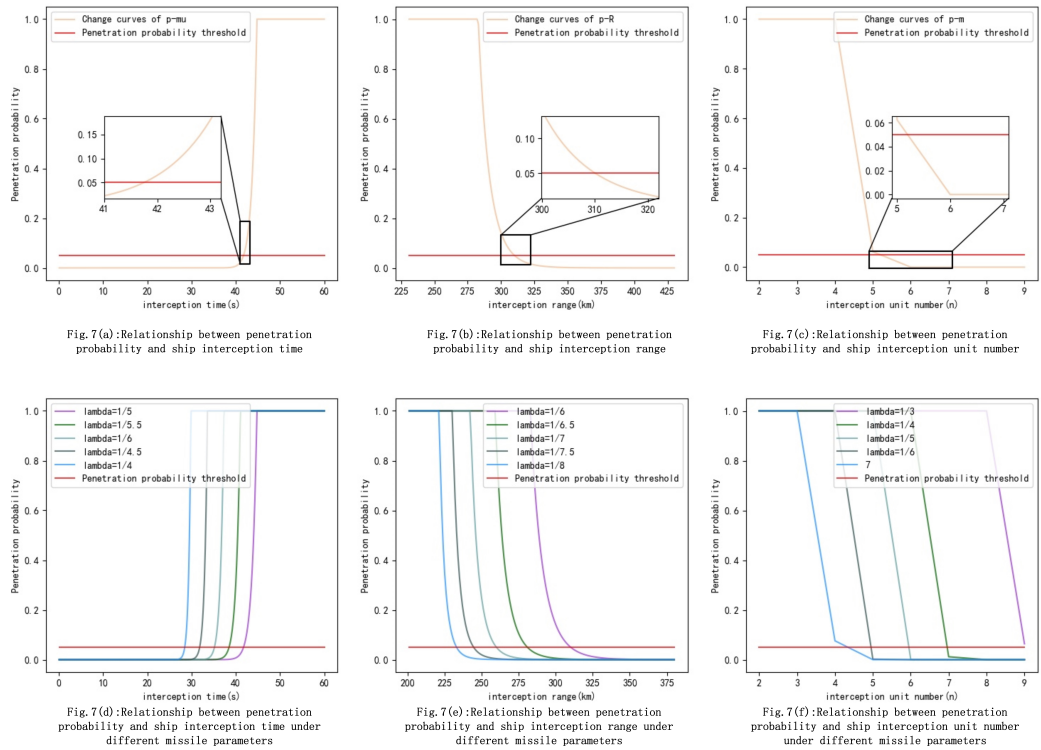}
 \caption{Relationship between ship parameters and penetration probability.}
 \label{fig:time}
\end{figure}

It can be seen from Fig. 7(a), Fig. 7(b), Fig. 7(c) that the penetration probability decreases with the increase of the interception rate, interception range and number of interception units of the ship. This conclusion is in line with the experimental expectations. The greater the interception rate, the less time it takes to intercept a missile, so the probability of penetration is lower; The larger the interception range, the longer the flight time of the missile from entering the interception range to hitting the target, that is, the longer the waiting time, the greater the penetration probability; The more the number of intercepting units, the more missiles each ship can intercept at the same time, so the probability of penetration decreases. Fig. 7(d), Fig. 7(e), Fig. 7(f) reflect that different model hyperparameters correspond to different curve clusters.

At the same time, it is noted that the range of penetration probability varies with the parameters, and the range of change with the interception rate is greater than that of the interception range. This result can be explained by the penetration probability expression. From the Eq. (26), the expression of the penetration probability about the interception rate $\mu$ is the Eq. (33), and the expression about the interception range R is the Eq. (34), where A is a constant. From the expression, it is known that the order of the penetration probability P with respect to the interception rate $\mu$ is higher, so the curve is steeper. Since the number of intercepting units can only be changed in integer type, it is a polyline in Fig.7(c). From the steep change curve of the penetration probability above, it can be seen that with the enhancement of ship parameters, the missile penetration probability continues to decrease, but the reduction process is not a linear change, but after reaching a certain criticality, it rapidly drops exponentially to within the penetration threshold.

\begin{equation}
P=f(\mu)={\frac{1}{(m-1)! (m-{\frac{\lambda}{\mu}}) \sum_{i=0}^m {\frac{({\frac{\lambda}{\mu}})^{i-m}}{i!}}+ {\frac{1}{\mu}}}} e^{{-W_t} (m \mu - \lambda)}
\end{equation}

\begin{equation}
P=g(R)=A e^{-{\frac{m \mu (1-\rho)}{v}} R}
\end{equation}

Therefore, in practical applications, as a defender, the penetration probability of incoming missiles can be effectively reduced by increasing the interception rate, interception range and the number of interception units on the ship. However, when increasing the interception rate and interception range, there is no need to increase the interception limitlessly, and it only needs to reach a certain critical value. After the critical value is exceeded, the cost increases and the penetration probability does not change significantly.

\subsubsection{The Relationship between Missile Parameters and Penetration Probability}

This part also uses the control variable method to explore the relationship between the missile parameters at the offensive end and the penetration probability. The experimental results are shown in Fig. 8. Fig. 8(a), Fig. 8(b), Fig. 8(c) are the change curves of the penetration probability when the missile launch rate, the missile flight speed and the missile launch direction change, respectively. Fig. 8(d), Fig. 8(e), and Fig. 8(f) are the curve clusters corresponding to the above curves when the parameters of the ship's interception range are changed.

\begin{figure}
 \centering
 \includegraphics[width=17cm]{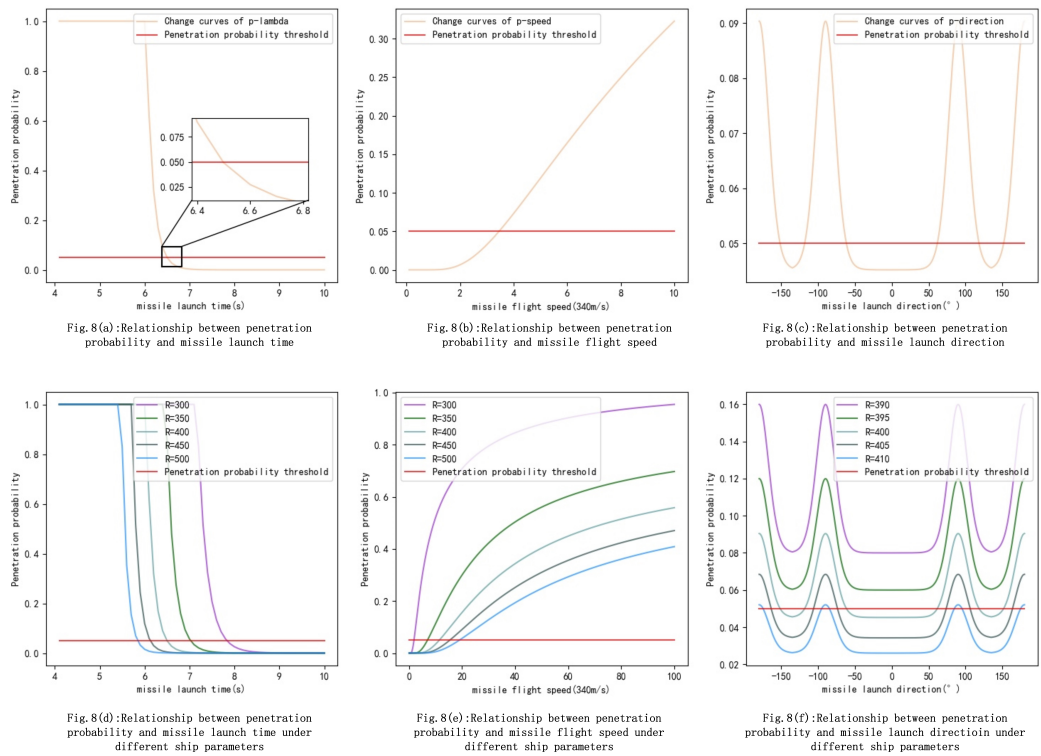}
 \caption{Relationship between offensive missile parameters and penetration probability.}
 \label{fig:time}
\end{figure}

It can be seen from Fig. 8(a), Fig. 8(b) that the penetration probability increases with the increase of the missile launch rate and flight speed. The conclusion is in line with the experimental expectations, the faster the launch frequency and flight speed of the incoming missile, the higher the penetration probability. But at the same time, it can be seen from the figure that the magnitude of the change in the penetration probability of the two is different. Fig. 8(a) shows that with the increase of the missile launch rate, the penetration probability does not change significantly at first, but it will suddenly increase exponentially after reaching a certain critical value, thus exceeding the penetration threshold to achieve penetration. Fig. 8(b) shows that as the flight speed of the missile increases, the penetration probability also increases steadily and synchronously. But at the same time, Fig. 8(e) shows that when the flight speed reaches a certain level, the increase rate of penetration probability will be greatly slowed down. In Fig. 8(e), when the flight speed increases from 20 times the speed of sound to 100 times the speed of sound, the penetration probability does not change significantly. The above experiments are theoretical experiments. In reality, the speed of missiles is far less than 20 times the speed of sound and above. 

Therefore, in practical applications, as an attacker, the missile penetration probability can be improved by increasing the missile launch rate and flight speed, but the ways of increasing the two are different. In reality, it is easy to increase the missile launch rate, but it only needs to reach a certain critical value when increasing. If the critical value is exceeded, the cost will increase and the penetration probability will not change much. In reality, it is difficult to increase the flight speed of the missile, because it involves technical bottlenecks, and the missile flight speed that can be achieved by the current technology is still far from the critical value when the change of the penetration probability in Fig. 8(e) slows down. Therefore, increasing the flight speed of missiles can effectively improve the probability of penetration, and the current development of supersonic missiles in various countries also confirms this.

It can be seen from Fig. 8(c) that the other three missiles are fixed, and when only the launch direction of one missile is changed, the penetration probability will change with the change of direction. Some directions have high penetration probability, and some directions have Low penetration probability. In this experiment, the direction of missile 1 is changed. The experimental results show that when missile 1 is launched from the direction of 90 degrees, -90 degrees, and ±180 degrees, the probability of model penetration is the largest. These three positions are exactly where the other three missiles were launched. The experimental results are in line with expectations, because when the missile 1 is launched in the same direction as other missiles, the load of the ship facing this direction will increase, and the ship will be more likely to be penetrated, so the penetration probability of the entire model will increase. Fig. 8(f) is the curve cluster of the penetration probability corresponding to different ship parameters with the direction. 

Therefore, n practical applications, how the attacker chooses the missile launch direction has a great impact on the penetration probability. In the case that the parameters of the attacker and the defender cannot be changed, the probability of penetration should be increased by analyzing the enemy's formation, identifying weak links, and reasonably selecting the missile launch position. The experimental results provide a way of thinking, that is, multiple missiles focus on one point and break through together, rather than spread out and strike widely.

\section{Conclusion}
Based on the background of maritime air defense and missile defense, this paper improves the traditional air defense model for the synergy between defense ships, and proposes a multi-ship cooperative air defense model. The parameters of this model are composed of the missile launch rate, flight speed and launch direction of the attacker, the ship formation arrangement, interception rate, interception range and number of interception units of the defender. The model can dynamically adjust the parameters according to the actual demand, with strong flexibility. This model assigns tasks to defense ships according to the principle of the proportion of missiles entering the interception range of ships. Different ships cooperate to intercept incoming missiles according to the task assignment results, and the overall penetration probability expression of the model is deduced by using queuing theory. Experiments show that under the same model parameters, the penetration probability of multi-ship cooperative interception of incoming missiles is smaller than that of each ship's isolated interception, which verifies the superiority of the multi-ship cooperative air defense model. Finally, this paper explores the significance of the model in practical application by analyzing the relationship between different parameters of the model and the probability of penetration. The multi-ship cooperative air defense model proposed in this paper has certain guiding significance for the research of air defense tactics, simulation actual combat deduction, weak link analysis and design of new weapons and equipment.

\bibliographystyle{unsrt}  
\bibliography{references}

\end{document}